# Solution of a Class of the Riemann-Papperitz Equation with Two Singular Points

Milan Batista

University of Ljubljana

Faculty of Maritime Studies and Transport

Portoroz, Slovenia, EU

milan.batista@fpp.edu

(Aug 10, 2005)

## Abstract

This paper provides the solution of the Riemann-Paperitz equation with singular points at $z = \pm \hat{i}$. This solution is obtained by mapping the singular points into points $0, \infty$. The solution is then obtained in terms of the Gauss hypergeometric function.

## 1  Introduction

The object is the following equation

$$\left(1+z^2\right)^2 \frac{d^2 y}{dz^2} + 2az\left(1+z^2\right)\frac{dy}{dz} + 4\left(b+cz\right) y = 0 \qquad (1)$$

where $y(z)$ is an unknown function, $z \in \mathbb{C}$ is a variable and $a, b, c \in \mathbb{C}$ are constants. (Factors 2 and 4 in (1) are introduced for convenience). The equation is a special case of the Riemann-Papperitz equation ([3],[4]) with only two regular singular points at $z = \pm\hat{i}$. The equation of type (1) appears in the dynamics of a disk (having finite thickness) rolling on a rough plane.

**Note**. The solution of the equation of type $(1+z^2)^2 \dfrac{d^2y}{dz^2} + az(1+z^2)\dfrac{dy}{dz} + by = 0$ can be found in [2] (equation 2.368)

The equation will be solved in the usual way ([3],[4]). First the equation's singular points will be mapped by bilinear transformations into the points $0, \infty$. Then the transformed equation will be reduced to the Gauss hypergeometric equation ([3],[4]). This equation has the form

$$t(1-t)\dfrac{d^2Y}{dt^2} + \left[\gamma - (\alpha + \beta + 1)t\right]\dfrac{dY}{dt} - \alpha\beta Y = 0 \qquad (2)$$

and has solutions including

$$Y = C_1 F(\alpha, \beta, \gamma, t) + C_2 t^{1-\gamma} F(\alpha - \gamma + 1, \beta - \gamma + 1, 2 - \gamma, t) \qquad (3)$$

where $F$ is the Gauss hypergeometric function with $\alpha, \beta, \gamma$ as parameters and $t$ as the argument, and $C_1, C_2$ are arbitrary constants.

## 2 Solution

The function

$$t = \dfrac{z - \hat{i}}{z + \hat{i}} \qquad (4)$$

where $\hat{i} = \sqrt{-1}$, maps singular points $z = -\hat{i}, \hat{i}$ to points $t = 0, \infty$. Also it maps the interior of the unit circle $|t| \leq 1$ to the upper half space $\Im z \geq 0$. The inverse transformation of (4) is

$$z = \hat{i}\dfrac{1+t}{1-t} \qquad (5)$$

Using (5), equation (1) is transformed to

$$t^2(1-t)\frac{d^2y}{dt^2}+t\left[a-(2-a)t\right]\frac{dy}{dt}+\left[(b-\hat{i}c)t-(b+\hat{i}c)\right]y=0 \qquad (6)$$

where $y(z(t)) = y(t)$. The solution of (6) is assumed to be of the form ([3])

$$y(t) = t^\lambda Y(t) \qquad (7)$$

where $\lambda$ is constant and $Y(t)$ the function which are both to be determined. Substituting (7) into (6) yields the hypergeometric equation

$$t(1-t)\frac{d^2Y}{dt^2}+\left[(2\lambda+a)-(2+2\lambda-a)t\right]\frac{dY}{dt}-\left[\lambda^2+(1-a)\lambda-(b-\hat{i}c)\right]Y=0 \qquad (8)$$

where $\lambda$ is any one of the solutions of the quadratic equation

$$\lambda^2 - (1-a)\lambda - (b+\hat{i}c) = 0 \qquad (9)$$

The solution $Y(t)$ of (8) has the form (3), where, by comparing (2) and (8), the parameters $\alpha, \beta, \gamma$ are

$$\gamma = 2\lambda + a \qquad \alpha + \beta = 1 + 2\lambda - a \qquad \alpha\beta = \lambda^2 + (1-a)\lambda - (b-\hat{i}c) \qquad (10)$$

It is seen from (10) that $\alpha$ and $\beta$ are solutions of the quadratic equation $z^2 - (\alpha+\beta)z + \alpha\beta = 0$. If the solution of (9) is selected to be

$$\lambda = \frac{1-a+\Delta}{2}, \qquad \Delta \equiv \sqrt{(1-a)^2 + 4(b+\hat{i}c)} \qquad (11)$$

then by (10), it follows that

$$\alpha = 1 - a + \frac{\Delta + \Delta^*}{2}, \qquad \beta = 1 - a + \frac{\Delta - \Delta^*}{2}, \qquad \gamma = 1 + \Delta$$
$$\Delta^* \equiv \sqrt{(1-a)^2 + 4(b - \hat{i}c)} \tag{12}$$

The solution of (1), using (7), (11) and (12), can now be written in the form

$$y = C_1 t^\lambda F(\alpha, \beta, \gamma, t) + C_2 t^{1+\lambda-\gamma} F(\alpha - \gamma + 1, \beta - \gamma + 1, 2 - \gamma, t) \tag{13}$$

or, reverting from $t$ to the variable $z$, using (4), of

$$y = C_1 \left(\frac{z - \hat{i}}{z + \hat{i}}\right)^\lambda F\left(\alpha, \beta, \gamma, \frac{z - \hat{i}}{z + \hat{i}}\right)$$
$$+ C_2 \left(\frac{z - \hat{i}}{z + \hat{i}}\right)^{1+\lambda-\gamma} F\left(\alpha - \gamma + 1, \beta - \gamma + 1, 2 - \gamma, \frac{z - \hat{i}}{z + \hat{i}}\right) \tag{14}$$

Solution (14) can be put in various forms by using the transforming formulas for $F$ functions ([1][3][4]).

## 3 Conclusion

A solution of the equation

$$(1 + z^2)^2 \frac{d^2 y}{dz^2} + 2az(1 + z^2)\frac{dy}{dz} + 4(b + cz) y = 0 \tag{15}$$

is

$$y = C_1 \left(\frac{z - \hat{i}}{z + \hat{i}}\right)^\lambda F\left(\alpha, \beta, \gamma, \frac{z - \hat{i}}{z + \hat{i}}\right)$$
$$+ C_2 \left(\frac{z - \hat{i}}{z + \hat{i}}\right)^{1+\lambda-\gamma} F\left(\alpha - \gamma + 1, \beta - \gamma + 1, 2 - \gamma, \frac{z - \hat{i}}{z + \hat{i}}\right) \tag{16}$$

where

$$\lambda = \frac{1-a+\Delta}{2}, \quad \alpha = 1-a+\frac{\Delta+\Delta^*}{2}, \quad \beta = 1-a+\frac{\Delta-\Delta^*}{2}, \quad \gamma = 1+\Delta \quad (17)$$

and

$$\Delta \equiv \sqrt{(1-a)^2 + 4(b+\hat{i}c)} \quad \Delta^* \equiv \sqrt{(1-a)^2 + 4(b-\hat{i}c)} \quad (18)$$